\documentclass[12pt,preprintnumbers,nofootinbib,aps,prd]{revtex4-1}

\usepackage[T1]{fontenc}
\usepackage{lmodern}

\usepackage{amsmath}
\usepackage{amssymb}
\usepackage{mathtools}
\usepackage{graphicx}
\usepackage{relsize}
\usepackage{exscale}
\usepackage{wasysym}
\usepackage{ulem}
\usepackage{cancel}
\usepackage{braket}
\usepackage{titlesec}
\usepackage{float}
\usepackage{enumitem}
\usepackage{tensor}

\usepackage{graphicx}
\usepackage{slashed}
\usepackage[left=1cm,right=1cm,top=2cm,bottom=2cm]{geometry}

\begin{document}

\preprint{ITP-UU-13/19, SPIN-13/12, UFIFT-QG-13-07}
\title{Electrodynamic Effects of Inflationary Gravitons}

\author{D.~Glavan}
\email[ \, ]{d.glavan@uu.nl}
\author{T.~Prokopec}
\email[ \, ]{t.prokopec@uu.nl}
\affiliation{Institute for Theoretical Physics \& Spinoza Institute \\
Utrecht University, Postbus 80195, 3508 TD Utrecht, THE NETHERLANDS}

\author{S.~P.~Miao}
\email[ \, ]{spmiao5@mail.ncku.edu.tw}\affiliation{Department of Physics, 
National Cheng Kung University \\ No.1, University Road, Tainan City 701, 
TAIWAN}

\author{R.~P.~Woodard}
\email[ \, ]{woodard@phys.ufl.edu}
\affiliation{Department of Physics, University of Florida,\\
Gainesville, 32611 FL, UNITED STATES}

\begin{abstract}
We calculate the one-loop corrections from inflationary gravitons to
the electromagnetic fields of a point charge and a point magnetic
dipole on a locally de Sitter space background. Results are obtained
both for an observer at rest in co-moving coordinates, whose
physical distance from the sources increases with the expanding
universe, and for an observer at rest in static coordinates, whose
physical distance from the sources is constant. The fields of both
sources show the de Sitter analogs of the fractional $G/r^2$
corrections which occur in flat space, but there are also some
fractional $G H^2$ corrections due to the scattering of virtual
photons from the vast ensemble of infrared gravitons produced by
inflation. The co-moving observer perceives the magnitude of the
point charge to increase linearly with co-moving time and
logarithmically with the co-moving position, however, the magnetic
dipole shows only a negative logarithmic spatial variation. The
static observer perceives no secular change of the point charge but
he does report a secular enhancement of the magnetic dipole moment.
\end{abstract}

\maketitle

\section{Introduction}

\begin{figure}[ht]
\hspace{-9cm} \includegraphics[width=4.0cm,height=3.0cm]{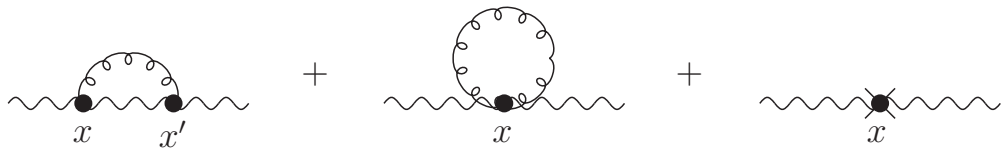}
\vspace{-2cm}
\caption{\tiny \label{fig:photon} Feynman diagrams relevant to the
one loop vacuum polarization from gravitons. Wavy lines are photons
and curly lines are gravitons.}
\end{figure}

Primordial inflation produces a vast ensemble of scalars and
gravitons which are the sources of primordial scalar and tensor
perturbations \cite{perts}. These ensembles can alter the properties
of particles and the forces they carry. Many studies of these
modifications have been made in recent years, both for
scalar-mediated effects \cite{phi4,SQED,Yukawa,PW,DW} and for
graviton-mediated effects \cite{MW1,SPM,KW1,HKYK,PMTW}. The aim of
this paper is to determine the leading corrections from inflationary
gravitons to the electric and magnetic fields produced by a point
charge and by a point magnetic dipole. Our technique is to solve the
quantum-corrected Maxwell's equation,
\begin{equation}
\partial_{\nu} \left[\sqrt{-g} g^{\nu\alpha} g^{\mu\beta}
F_{\alpha\beta}(x) \right] + \int\! d^4x'
\big[{}^\mu{\Pi}_R^\nu\big](x;x')\, A_{\nu}(x') = J^{\mu}(x) \ ,
\label{genEFE}
\end{equation}
where $F_{\mu\nu} = \partial_{\mu} A_{\nu} - \partial_{\nu} A_{\mu}$
is the electromagnetic field strength tensor, $A_{\mu}$ is the
electromagnetic 4-potential, $J^{\mu}$ is the source 4-current,
$i[{}^\mu\Pi_R^\nu](x;x')$ is the retarded vacuum polarization
induced by the interactions with gravitons. We infer the retarded
vacuum polarization using the Schwinger-Keldysh formalism \cite{SK}
from a recent computation of the one loop graviton contribution to
the vacuum polarization on de Sitter background, made using
dimensional regularization and BPHZ
(Bogoliubov-Parasiuk-Hepp-Zimmermann) renormalization \cite{LW1}.
The relevant diagrams are displayed in Fig.~\ref{fig:photon}.
\vspace{1cm} 

Our analysis is highly relevant to three earlier works
\cite{DW,HKYK,LW2}. The last of these is a study of graviton
corrections to electromagnetism on flat space background. Because
the same sources were included, this work gives the flat space
correspondence limits of our de Sitter results. Static fields in
flat space can only depend upon the distance $r$ from the source, so
one loop quantum gravitational corrections must be proportional to
the classical result times $G/r^2$, where $G$ is Newton's constant.
Of course explicit computation \cite{LW2,BB} confirms this simple
consequence of dimensional analysis.

Our de Sitter problem has another dimensional parameter in the form
of the Hubble constant $H$. It can also show secular growth,
deriving ultimately from the fact that more and more gravitons are
ripped out of the vacuum as inflation progresses. These features
mean the classical fields can suffer fractional corrections of the
form $G H^2 \times \ln(a)$, where $a(t) = e^{H t}$ is the de Sitter
scale factor. Because those corrections have the same spatial
dependence as the classical result it seems fair to regard them as
time dependent renormalizations of the classical sources, which in
our case are the charge and the magnetic dipole moment. Precisely
this sort of secular renormalization was seen in a study of the
effect of charged inflationary scalars on the same two sources
\cite{DW}. Because that study found different results for observers
at a fixed physical distance from the source and those who are being
pulled away by the inflationary expansion, we shall also derive
results for both cases. And a major motivation for our work is to
check the recent claim by Kitamoto and Kitazawa that inflationary
gravitons screen gauge coupling constants \cite{HKYK}.

Section II of this paper recasts the results of Ref.~\cite{LW1} in
Schwinger-Keldysh form \cite{SK} to give the one loop retarded
vacuum polarization. In section III we make a loop expansion on the
field strength of the effective field and derive an integral
expression for the one loop contribution in terms of the tree order
field strengths and the structure functions of the vacuum
polarization. Section IV gives the actual derivation of the field
strengths for a point charge and for a point magnetic dipole, with
some technical details consigned to appendices. Our conclusions
comprise section V.

\section{The Retarded Vacuum Polarization}

The de Sitter metric tensor in spatially flat, conformal coordinates
is $g_{\mu\nu}=a^2(\eta) \eta_{\mu\nu}$. Here and henceforth, the
Minkowski metric is $\eta_{\mu\nu}=\mathrm{diag}(-1,1,1,1)$, and
$a(\eta)=-{1}/(H\eta)$ is the scale factor in terms of conformal
time $\eta$. Because the vacuum polarization is a transverse
bi-vector density it can be written in the form \cite{SQED},
\begin{equation} \label{generalVacPol}
i \big[{}^\mu\Pi^\nu\big](x;x')
  = (\eta^{\mu\nu}\eta^{\rho\sigma} - \eta^{\mu\sigma}\eta^{\nu\rho})
\partial_{\rho} \partial_{\sigma'} F(x;x')
+ (\overline{\eta}^{\mu\nu} \overline{\eta}^{\rho\sigma}
- \overline{\eta}^{\mu\sigma} \overline{\eta}^{\nu\rho}) \partial_{\rho}
            \partial_{\sigma'} G(x;x') \ ,
\end{equation}
Here and henceforth, placing a bar over a tensor indicates that its
temporal components have been suppressed, for example,
$\overline{\eta}^{\mu\nu} \equiv \eta^{\mu\nu} + \delta_{0}^{\mu}
\delta_{0}^{\nu}$. $F(x;x')$ and $G(x;x')$ are known as {\it
structure functions}, and one can show that two of them are needed
if the only coordinate symmetries are homogeneity and isotropy
\cite{LPW1}. Because the graviton propagator breaks de
Sitter invariance \cite{TW1,Kleppe}, a de Sitter breaking
representation like (\ref{generalVacPol}) is mandatory. Had all ten
of the de Sitter isometries been present one could employ a hugely
more complicated but de Sitter invariant representation involving
only a single structure function. However, this representation seems
to obscure, rather than elucidate, the essential physics
\cite{LPW1}. And a simple procedure exists for transforming between
different representations \cite{LPW2}.

The one loop graviton contributions to the structure functions of
the renormalized, in-out vacuum polarization were given in
Eqs.~(136-137) of Ref.~\cite{LW1}. After some straightforward
rearrangements, those results can be expressed as,
\begin{eqnarray}
\lefteqn{F(x;x') = \frac{\kappa^2}{8\pi^2} \left\{ H^2 \left[ \ln
(a) \!+\! \alpha \right] \!+\! \frac{1}{a^2}\Big[-\frac{1}{3}\ln(a)
\!+\! \beta \Big] \left[\partial^2 \!+\! 2Ha\partial_0 \right]
\!+\! \frac{H}{3a} \partial_0 \right\} i \delta^4(x\!-\!x') } \nonumber \\
& & \hspace{.5cm} - \frac{\kappa^2}{1536\pi^4} \frac{1}{a}
\partial^6 \left\{ \frac{1}{a'} \left[ \ln^2\left(
\tfrac{H^2}{4}\Delta x^2 \right) - 2\ln\left(
\tfrac{H^2}{4}\Delta x^2 \right) \right] \right\} \nonumber \\
& & \hspace{1cm} + \frac{\kappa^2H^2}{128\pi^4} \left\{ \Big[
\frac{1}{4}\partial^4 \!+\! \partial^2\partial_0^2 \Big] \ln^2\left(
\tfrac{1}{4} H^2 \Delta x^2 \right) + \Big[ \!-\! \frac{1}{2}
\partial^4 \!+\! 2 \partial^2\partial_0^2 \Big] \ln\left(
\tfrac{1}{4} H^2 \Delta x^2 \right) \right\} + O(\kappa^4) \; ,
\qquad \label{Fdef} \\
\lefteqn{G(x;x') = \frac{\kappa^2H^2}{6\pi^2} \left[-\ln(a) \!+\!
\frac{3}{4} \gamma \right] i \delta^4 (x\!-\!x') } \nonumber \\
& & \hspace{5.5cm} - \frac{\kappa^2H^2}{384 \pi^4} \partial^4
\Bigl\{ \ln^2\left( \tfrac{1}{4} H^2 \Delta x^{2} \right) \!-\!
2\ln\left( \tfrac{1}{4} H^2 \Delta x^{2} \right) \Bigr\} +
O(\kappa^4) \; . \label{Gdef} \qquad
\end{eqnarray}
Here the loop counting parameter of quantum gravity is
$\kappa^2=16\pi G$, the flat space d'Alembertian is $\partial^2
\equiv \eta^{\mu\nu} \partial_{\mu} \partial_{\nu}$ and we define
the invariant interval $\Delta x^2(x;x')$ as,
\begin{equation}
\Delta x^2 = - (|\eta\!-\!\eta'|\!-\!i\epsilon)^2
      \!+\!\| \vec{x}\!-\!\vec{x}^{\,\prime}\|^2 \; .
\end{equation}
The parameters $\alpha$, $\beta$ and $\gamma$ in
(\ref{Fdef}-\ref{Gdef}) are finite renormalization constants that
are related to the one loop counterterms used in Ref.~\cite{LW1},
\begin{eqnarray}
{\cal L}_{\rm ct}& = & C_1\sqrt{-g}RF_{\mu\nu}F^{\mu\nu}
+C_2\sqrt{-g}R^{\nu\sigma}F_{\mu\nu}F_{\rho\sigma}g^{\mu\rho}
+C_3\sqrt{-g}R^{\mu\nu\rho\sigma}F_{\mu\nu}F_{\rho\sigma}
\nonumber\\
& & +C_4\sqrt{-g}(\nabla_\alpha F_{\mu\nu})(\nabla^\alpha
F^{\mu\nu}) +\Delta C H^2\sqrt{-g}F_{ij}F_{kl}g^{ik}g^{jl} \, .
\quad \label{counterterms}
\end{eqnarray}
In de Sitter background only the combination $\overline{C} =
D(D\!-\!1) C_1 + (D\!-\!1) C_2 + 2 C_3$ matters. We define $C_{4f}$,
$\overline{C}_f$ and $\Delta C_f$ as the finite parts of each
coefficient, and the parameters $\alpha$, $\beta$ and $\gamma$ are,
\begin{eqnarray}
\alpha =  \frac{1}{2} \bigg[3\ln\bigg(\frac{4\mu^2}{H^2}\bigg) - 4
\bigg] + \frac{32\pi^2}{\kappa^2} (\overline{C}_f - 4 C_{4f}) \quad
, \quad \beta = -\frac{1}{6} \ln\bigg(\frac{4\mu^2}{H^2}\bigg) -
\frac{32\pi^2}{\kappa^2}C_{4f} \; , \nonumber \\
\gamma = \frac{1}{3} \bigg[\ln\bigg(\frac{4\mu^2}{H^2}\bigg) -
\frac{33}{2}\bigg] + \frac{32\pi^2}{\kappa^2}(\Delta C_f - 2 C_{4f})
\, . \label{cparams}
\end{eqnarray}
Had Einstein + Maxwell been a renormalizable theory we could have
invoked some physical renormalization condition to fix the 
coefficients $\alpha$, $\beta$ and $\gamma$. However, Einstein +
Maxwell is not perturbatively renormalizable \cite{DvN}, so we must
instead treat it in the sense of low energy effective field theory
\cite{Donoghue}. That is, we regard the finite renormalization
constants $\alpha$, $\beta$ and $\gamma$ as arbitrary free 
parameters which characterize our ignorance of the true ultraviolet
completion of Einstein + Maxwell, and we trust only predictions of
the theory which are insensitive to the values of these parameters.
Precisely this was done by Bjerrum-Bohr in his computation of the
long range one loop graviton contribution to the Coulomb potential 
on a flat background \cite{BB}. We will comment further on this at
the appropriate points of subsections \ref{Point charge} and 
\ref{dipole}, after our full results for the one loop field strengths 
have been derived and it is possible to identify regimes in which 
the unambiguous contributions dominate those from $\alpha$, $\beta$ 
and $\gamma$.

Using the in-out structure functions (\ref{Fdef}-\ref{Gdef}) in
Eq.~(\ref{genEFE}) would be appropriate for a flat space scattering
problem but it makes little sense in cosmology where the universe
began with an initial singularity and no one knows its final state.
Using the in-out structure functions would make the effective field
equations depend strongly on the far future; it would also result in
the electromagnetic field strengths developing imaginary parts. The
more appropriate problem to study in cosmology is what happens to
the field strengths when the universe is released in a prepared
state at some finite time. The appropriate structure functions for
this sort of problem are the retarded ones of the Schwinger-Keldysh
formalsim \cite{SK}.

Fortunately, there is a very simple procedure for converting in-out
structure functions into retarded ones \cite{FW},
\begin{equation}
F_R(x;x^\prime) = F_{\scriptscriptstyle ++}(x;x^\prime) +
F_{\scriptscriptstyle +-}(x;x^\prime) \, , \qquad
G_R(x;x^\prime) = G_{\scriptscriptstyle ++}(x;x^\prime) +
G_{\scriptscriptstyle +-}(x;x^\prime) \label{retarded F and G:def}
\end{equation}
We extract the $++$ and $+-$ structure functions from
(\ref{Fdef}-\ref{Gdef}) by replacing the invariant interval $\Delta
x^2(x;x')$ with,
\begin{equation}
\Delta x^2_{\scriptscriptstyle ++}(x;x') \equiv -(\vert \eta \!-\!
\eta' \vert \!-\! i \epsilon)^2 + \Vert \vec{x} \!-\! \vec{x}'
\Vert^2 \qquad , \qquad \Delta x^2_{\scriptscriptstyle +-}(x;x')
\equiv -(\eta \!-\! \eta' \!+\! i \epsilon)^2 + \Vert \vec{x} \!-\!
\vec{x}' \Vert^2 \; .
\end{equation}
We also drop the delta function terms in the $+-$ case, and
introduce an overall minus sign. The result is,
\begin{align}
-i F_R(x;x')
={}& \frac{\kappa^2}{8\pi^2} \left\{ H^2
       \left[ \ln(a)\!+\!\alpha \right]
   + \frac{1}{a^2} \Big[\!-\!\frac{1}{3}\ln(a) + \beta \Big]
\left[ \partial^2 + 2Ha\partial_0 \right]
+ \frac{H}{3a} \partial_0 \right\} \delta^4(x\!-\!x^{\,\prime})
\nonumber \\
& -\frac{\kappa^2}{384\pi^3} \frac{1}{a} \partial^6
\left\{\frac{1}{a'}\theta(\Delta\eta\!-\!\|\vec{x}\!-\!\vec{x}^{\,\prime}\|^2)
\left[ \ln\left[\tfrac{H^2}{4}
      \left( \Delta\eta^2\!-\!\| \vec{x}\!-\!\vec{x}^{\,\prime}\|^2
\right) \right] \!-\! 1 \right] \right\} \nonumber \\
& + \frac{\kappa^2H^2}{32\pi^3}
 \Big{\{} \Big[\frac{1}{4}\partial^4 \!+\! \partial^2\partial_0^2 \Big]
       \theta(\Delta\eta\!-\! \| \vec{x}\!-\!\vec{x}^{\,\prime}\|)
 \ln\left[\tfrac{H^2}{4}
         \left(\Delta\eta^2\!-\!\|\vec{x}\!-\!\vec{x}^{\,\prime}\|^2
    \right) \right] \nonumber \\
& \ \ \ \ \ \ \ \ \ \
+\Big[\!-\frac{1}{4}\partial^4\!+\!\partial^2\partial_0^2
   \Big] \theta(\Delta\eta\!-\!\|\vec{x}\!-\!\vec{x}^{\,\prime}\|)\Big\}
+ O(\kappa^4) \ , \label{FR} \\
-i G_R(x;x')={}& \frac{\kappa^2H^2}{8\pi^2}
         \Big[\!-\!\frac{4}{3}\ln(a)\!+\!\gamma \Big]
                \delta^4(x\!-\!x^{\,\prime})
\nonumber \\
& - \frac{\kappa^2 H^2}{96\pi^3} \partial^4
     \left\{ \theta(\Delta\eta\!-\! \| \vec{x}\!-\!\vec{x}^{\,\prime}\|)
     \left[\ln\left[ \tfrac{H^2}{4}
      \left( \Delta\eta^2 \!-\!\| \vec{x}\!-\!\vec{x}^{\,\prime}\|^2 \right)
\right] \!-\! 1\right] \right\} + O(\kappa^4) \ . \label{GR}
\end{align}
To reach these forms we have used the identities,
\begin{align}
& \ln\left( \tfrac{H^2}{4} \Delta x_{++}^2 \right) - \ln\left(
\tfrac{H^2}{4} \Delta x_{+-}^2 \right) = 2i\pi \,
\theta(\Delta\eta\!-\!\|\vec{x}\!-\!\vec{x}^{\,\prime}\|) \ ,
\label{imaginary logs} \\
& \ln^2\left( \tfrac{H^2}{4} \Delta x_{++}^2 \right) - \ln^2\left(
\tfrac{H^2}{4} \Delta x_{+-}^2 \right) = 4i\pi \,
\theta(\Delta\eta\!-\!\|\vec{x}\!-\!\vec{x}^{\,\prime}\|) \ln\left[
\tfrac{H^2}{4} \left[\Delta\eta^2\!-\!\|
\vec{x}\!-\!\vec{x}^{\,\prime}\|^2 \right] \right] \,. \quad
\label{imlogs2}
\end{align}
The retarded structure functions $-i F_R(x;x')$ and $-i G_R(x;x')$
in~(\ref{FR}--\ref{GR}) and the corresponding vacuum polarization
tensor are manifestly real and causal (in the sense that they vanish
outside the past light-cone).

These rules correspond to releasing the universe in free vacuum. If
the initial state has corrections --- as it must when interactions
are present --- there will be interactions on the initial value
surface \cite{KOW}. With the simple representation
(\ref{generalVacPol}) we are using, these temporal surface terms
should fall off like powers of $1/a$ \cite{LPW1}. Because we are
only interested in the asymptotic late time forms of the
quantum-corrected field strengths we will not bother correcting the
initial state.

\section{One Loop Effective Field Equation}

We can rewrite the effective field equation~\eqref{genEFE} in a
convenient form by noting that all the scale factors cancel in
$\sqrt{-g} \, g^{\nu\alpha} g^{\mu\beta} = \eta^{\nu\alpha}
\eta^{\mu\beta}$, by plugging in the vacuum
polarization~\eqref{generalVacPol}, and by partially integrating the
primed derivatives,
\begin{equation}\label{EFEreorg}
\partial_{\nu} F^{\nu\mu}(x) = J^{\mu}(x) - \partial_{\nu} \int \!
d^4x' \Big{[} i F_R(x;x') F^{\nu\mu}(x')
  + iG_R(x;x') \overline{F}^{\nu\mu} \Big{]} \; .
\end{equation}
Here and henceforth the indices on the field strength tensor are
raised with the Minkowski metric, $F^{\mu\nu} = \eta^{\mu\alpha}
\eta^{\nu\beta} F_{\alpha\beta}$, and we remind that reader that an
overline indicates the tensor has its temporal components
suppressed, $\overline{F}^{\mu\nu} \equiv
\overline{\eta}^{\mu\alpha} \overline{\eta}^{\nu\beta}
F_{\alpha\beta}$, with $\overline{\eta}^{\mu\nu} \equiv
\eta^{\mu\nu} + \delta^{\mu}_0 \delta^{\nu}_0$.

Because we only know the structure functions at order $\kappa^2$
there is no alternative to solving (\ref{EFEreorg}) in the loop
expansion,
\begin{eqnarray}
F^{\mu\nu}(x) & = & F^{\mu\nu}_{(0)}(x) + \kappa^2
F^{\mu\nu}_{(1)}(x) + \mathcal{O}(\kappa^4) \; , \\
F_R(x;x') & = & 0 + \kappa^2 F_{(1)}(x;x') + \mathcal{O}(\kappa^4)
\; , \\
G_R(x;x') & = & 0 + \kappa^2 G_{(1)}(x;x') + \mathcal{O}(\kappa^4)
\; .
\end{eqnarray}
We assume the current density $J^{\mu}(x)$ is classical, so the
zeroth order and the first order equations are,
\begin{align}
\partial_{\nu} F^{\nu\mu}_{(0)}(x) ={}& J^{\mu}(x)
\ ,
\label{F0mn}
\\
\partial_{\nu} F^{\nu\mu}_{(1)}(x)
 ={}& \partial_{\nu} \!\! \int \! d^4x'
 \left[ -iF_{(1)}(x;x') F^{\nu\mu}_{(0)}(x')
       -iG_{(1)}(x;x') \overline{F}^{\nu\mu}_{(0)}(x') \right] \ .
\label{F1mn}
\end{align}
Given the source $J^{\mu}(x)$ one solves equation (\ref{F0mn}) to
find the classical field strengths $F^{\nu\mu}_{(0)}(x)$, then one
uses this in (\ref{F1mn}), together with the order $\kappa^2$ parts
of (\ref{FR}-\ref{GR}), to solve for the one loop field strengths.

An important intermediate step is working out the primed integral on
the right hand side of ({\ref{F1mn}), without the unprimed
derivative,
\begin{equation}
\mathcal{F}^{\mu\nu}(x) \equiv \int \! d^4x' \left[ -iF_{(1)}(x;x')
F^{\mu\nu}_{(0)}(x') -iG_{(1)}(x;x') \overline{F}^{\mu\nu}_{(0)}(x')
\right] \; , \label{F1integral}
\end{equation}
where we assume the initial time is $\eta_0 = -1/H$. (The Heaviside
theta functions in Eqs.~(\ref{FR}--\ref{imlogs2}) then
dictate that the integration over $\eta^\prime$ is from $-{1}/{H}$
to $\eta$.) The divergence of $\mathcal{F}^{\mu\nu}$ defines what we
might call the one loop current $J^{\mu}_{(1)}$, which sources the
one loop field strength,
\begin{equation}
J^{\mu}_{(1)}(x) \equiv \partial_{\rho} \mathcal{F}^{\rho\mu}(x)
\qquad \Longrightarrow \qquad \partial^2 F^{\mu\nu}_{(1)}(x) =
\partial^{\mu} J^{\nu}_{(1)}(x) - \partial^{\nu} J^{\mu}_{(1)}(x) \;
. \label{1loopeqn}
\end{equation}
Because the retarded Green's function of $\partial^2$, $-\delta(\eta
- \eta' - \Vert \vec{x} - \vec{x}' \Vert)/4\pi \Vert \vec{x} -
\vec{x}' \Vert$, is translation invariant in space and time, we can
partially integrate the primed derivatives\footnote{\tiny Causality
precludes spatial surface terms in the Schwinger-Keldysh formalism
\cite{SK}. Nor can there be any future temporal surface terms, but
there can be --- and are --- initial time surface terms which we
ignore. These can perhaps be absorbed into perturbative corrections
of the initial state \cite{KOW} and probably fall off like powers of
$1/a$ \cite{LW1}.} and then reflect them to unprimed derivatives to
reach the form,
\begin{equation}
F^{\mu\nu}_{(1)} = \partial^{\mu} \partial_{\rho} \frac1{\partial^2}
\mathcal{F}^{\rho\nu} - \partial^{\nu}
\partial_{\rho} \frac1{\partial^2} \mathcal{F}^{\rho\mu} \; .
\label{basiceqn}
\end{equation}

An important simplification occurs when the classical field
strengths derive from static scalar or vector potentials. In that
case one can partially integrate the spatial derivatives onto the
one loop structure functions $F_{(1)}(x;x')$ and $G_{(1)}(x;x')$,
and then exploit homogeneity to reflect them into unprimed
derivatives. For the case of a static scalar potential, with zero
vector potential, we find,
\begin{equation}
A^{\mu}_{(0)}(\eta,\vec{x}) = \Bigl(-\Phi_{(0)}(\vec{x}),
\vec{0}\Bigr) \;\; \Longrightarrow \;\; \mathcal{F}^{0i} =
-\partial_i \!\! \int \!\! d^4x' \, i F_{(1)}(x;x')
\Phi_{(0)}(\vec{x}') \equiv -\partial_i \mathcal{J}_{\Phi} \;\; ,
\;\; \mathcal{F}^{ij} = 0 \; . \label{staticPhi1}
\end{equation}
Substituting (\ref{staticPhi1}) into (\ref{basiceqn}) tells us that
the one loop field strengths agree with $\mathcal{F}^{\mu\nu}$,
\begin{eqnarray}
\lefteqn{A^{\mu}_{(0)}(\eta,\vec{x}) = \Bigl(-\Phi_{(0)}(\vec{x}),
\vec{0}\Bigr) } \nonumber \\
& & \hspace{.8cm} \Longrightarrow  F^{0i}_{(1)} = \partial^0
\partial_0 \frac1{\partial^2} \, \mathcal{F}^{0i} \!-\! \partial^i
\partial_j \frac1{\partial^2} \, \mathcal{F}^{j0} = -\partial_i
\Bigl[ -\partial_0^2 \!+\! \nabla^2 \Bigr] \frac1{\partial^2} \,
\mathcal{J}_{\Phi} = \mathcal{F}^{0i} \quad , \quad F^{ij}_{(1)} = 0
\; . \qquad \label{staticPhi2}
\end{eqnarray}
Assuming a static and transverse vector potential yields,
\begin{equation}
A^{\mu}_{(0)}(\eta,\vec{x}) = \Bigl(0, \vec{m} \!\times\!
\vec{\nabla} \mathcal{A}(\vec{x}) \Bigr) \qquad \Longrightarrow
\qquad \mathcal{F}^{0i} = 0 \;\; , \;\; \mathcal{F}^{ij} =
\epsilon^{ijk} \Bigl[m_k \nabla^2 \!-\! \vec{m} \!\cdot\!
\vec{\nabla} \partial_k\Bigr] \mathcal{J}_{\mathcal{A}} \; ,
\label{staticA1}
\end{equation}
where we define,
\begin{equation}
\mathcal{J}_{\mathcal{A}}(x) \equiv - \!\! \int \!\! d^4x' \, i
\Bigl[ F_{(1)}(x;x') \!+\! G_{(1)}(x;x') \Bigr]
\mathcal{A}(\vec{x}') \; . \label{Jdef}
\end{equation}
Substituting (\ref{staticA1}) into (\ref{basiceqn}) allows us to
express the one loop field strengths in terms of
$\nabla^2/\partial^2 \, \mathcal{J}_{\mathcal{A}}$,
\begin{eqnarray}
\lefteqn{A^{\mu}_{(0)}(\eta,\vec{x}) = \Bigl(0, \vec{m} \!\times\!
\vec{\nabla} \mathcal{A}(\vec{x}) \Bigr) } \nonumber \\
& & \hspace{2cm} \Longrightarrow F^{0i}_{(1)} = -\epsilon^{ijk} m_j
\partial_k \partial_0 \Bigl( \frac{\nabla^2}{\partial^2} \,
\mathcal{J}_{\mathcal{A}} \Bigr) \;\; , \;\; F^{ij}_{(1)} =
\epsilon^{ijk} \Bigl[m_k \nabla^2 \!-\! \vec{m} \!\cdot\!
\vec{\nabla} \partial_k\Bigr] \Bigl( \frac{\nabla^2}{\partial^2} \,
\mathcal{J}_{\mathcal{A}} \Bigr) \; . \qquad \label{staticA2}
\end{eqnarray}

\section{The one-loop correction to the field strengths}
\label{The one-loop correction to the field strengths}

The purpose of this section is to derive the one loop field
strengths for a point charge (sub-section \ref{Point charge}) and
for a point magnetic dipole (sub-section \ref{dipole}). In each case
the computation is first made with the source at the origin in flat,
conformal coordinates (see Fig.~\ref{condiag}) and the observer at
fixed $x \equiv \Vert \vec{x} \Vert$, which gives the result for an
observer who is being
\begin{figure}[ht]
\vskip -2.2cm
\includegraphics[scale=0.9]{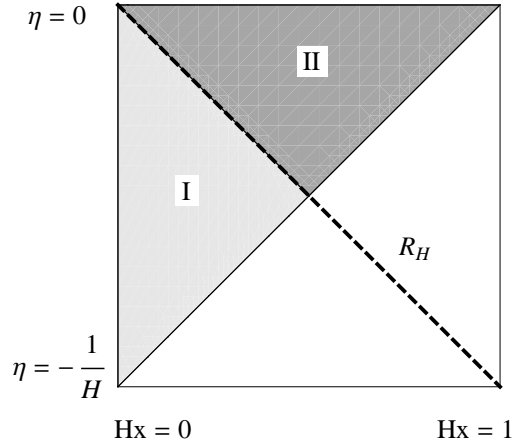}
\vskip -2.2cm

\caption{\tiny The conformal diagram of de Sitter space. The
universe is released at $\eta_0 = -1/H$ and we show only the region
$\eta_0 < \eta < 0$ and $0 \leq x \leq 1/H$. Our point sources are
at $x = 0$. Regions~I and~II are in causal contact with the initial
instant of the sources. The dashed line at $x = -\eta$, corresponds
to the physical distance (at fixed $\eta$) from the source being one
Hubble radius $R_H=1/H$. This line splits the causal region into
Part~I (sub-Hubble) and Part~II (super-Hubble).}

\label{condiag}

\end{figure}
pulled apart from the source by the expansion of the universe. We
then transform to the frame of an observer at fixed physical
distance from the source, according to the formulae given in
Appendix A.

\subsection{Point charge}\label{Point charge}

The classical current of a static, point charge $q$ is
\begin{equation}
J^0(\eta,\vec{x}) = q \delta^3(\vec{x}) \ , \ \ \
J^i(\eta,\vec{x})=0 \; .
\end{equation}
Solving~(\ref{F0mn}) gives the corresponding classical field
strengths,
\begin{equation}
E^i_{(0)}\equiv F^{i0}_{(0)}(\eta,\vec{x}) = -\partial_i
\Bigl(\frac{q}{4 \pi x} \Bigr) \ , \ \ \ F^{ij}_{(0)}(\eta,\vec{x})
= 0 \qquad (x=\|\vec{x}\|) \ . \label{E:classical}
\end{equation}
Of course this corresponds to a static scalar potential of the type
just considered. Substituting $\Phi_{(0)} = q/4\pi x$ in relations
(\ref{staticPhi1}-\ref{staticPhi2}) allows us to express the nonzero
one loop field strength in terms of derivatives of integrals
$\mathcal{I}_{ab}$ which are given in Appendix B,
\begin{align}
\kappa^2 F^{0i}_{(1)}(x) ={}& \frac{q}{4\pi} \partial_i \bigg{\{}
\frac{\kappa^2H^2}{8\pi^2x} \left[ \ln(a) + \alpha \right] -
\frac{\kappa^2}{384\pi^3a} \partial^6
(\mathcal{I}_{22}-\mathcal{I}_{21})
\nonumber \\
&+ \frac{\kappa^2H^2}{32\pi^3}
  \left[\Big(\frac{1}{4}\partial^4 \!+\! \partial_0^2\partial^2 \Big)
  \mathcal{I}_{12}
+ \Big(\!-\!\frac{1}{4}\partial^4\!+\!\partial_0^2\partial^2\Big)\mathcal{I}_{11}
  \right] \bigg{\}}
\ .
\end{align}
Their derivatives are evaluated in Appendix C to produce the result,
\begin{align}
\kappa^2F_{(1)}^{0i}(\eta,\vec{x}) ={}& \frac{q}{4\pi} \partial_i
\bigg{\{}
  \theta(\Delta\eta_0\!-\!x) \frac{\kappa^2H^2}{8\pi^2 x}
\left[ \frac{1}{3a^2H^2x^2} \!+\! \ln(aHx) \!+\! \alpha
 \!-\! 2\ln\left(\frac{1-a^{-1}+Hx}{1-a^{-1}-Hx}\right) \right]
\nonumber \\
& + \theta(x\!-\!\Delta\eta_0) \frac{\kappa^2H^2}{8\pi^2x}
 \left[ \ln(a) \!+\! \alpha \!-\! 4 \!-\!\frac{1}{3(a\!-\!1)}
        \!-\! \frac{a}{3(a\!-\!1)^2}
 \!-\! 3\ln\left( 1\!-\!\frac{1}{a} \right) \right] \bigg{\}}
\ ,
\label{F1i0}
\end{align}
where $\Delta\eta_0=\eta\!-\!\eta_0=\eta\!+\!1/H$ and
$x=\|\vec{x}\,\|$.

If we think of the full field strength as the curl of a covariant
4-vector potential $A_{\mu} \equiv (\Phi,\vec{A})$, then expression
(\ref{F1i0}) can be viewed as a one loop correction to the Coulomb
potential,
\begin{eqnarray}
\lefteqn{\Phi_{(1)} = \Phi_{(0)} \times \frac{\kappa^2 H^2}{8 \pi^2}
\Biggl\{ \theta(\Delta\eta_0\!-\!x) \left[ \frac{1}{3a^2H^2x^2} +
\ln(aHx) + \alpha - 2\ln\left(\frac{1-a^{-1}+Hx}{1-a^{-1}-Hx}\right)
\right] } \nonumber \\
& & \hspace{3.5cm} + \theta(x\!-\!\Delta\eta_0)
 \left[ \ln(a) \!+\! \alpha \!-\! 4 \!-\!\frac{1}{3(a\!-\!1)}
        \!-\! \frac{a}{3(a\!-\!1)^2}
 \!-\! 3\ln\left( 1\!-\!\frac{1}{a} \right) \right] \Biggr\}
\ . \qquad \label{potentials:ratio charge}
\end{eqnarray}
A number of points about this result deserve comment. First, note
that nothing special happens at the Hubble radius, $x = -\eta$,
which is the dashed line in Fig.~\ref{condiag}. Second, it would be
a mistake to pay much attention to the branch of
(\ref{potentials:ratio charge}) with $x > \Delta \eta_0$. As one can
see from Fig.~\ref{condiag}, observers in this region are not in
causal contact with the point source; they only feel its influence
as a consequence of whatever assumption is made about the long range
fields which are present in the initial state. Understanding of this
issue is in its infancy \cite{KOW}. The need for perturbative
corrections to the initial state is obvious from the constraint
equations, and from the singularities which occur at $\eta = \eta_0$
on the last line of (\ref{potentials:ratio charge}), but no one has
worked out these corrections. The same comments apply to the
logarithmic singularity on the first line of (\ref{potentials:ratio
charge}) which propagates along the light-cone from the initial
appearance of the point source at $x^{\mu} = (\eta_0,\vec{0})$. Note
that all of these terms fall off at late time like powers of $1/a$,
which marks them as artifacts of the initial state.

The factor of $\alpha$ on the first line of (\ref{potentials:ratio
charge}) is also unimportant. It depends on whatever assumption we
make about the finite parts of the counterterms
(\ref{counterterms}-\ref{cparams}). No physical principle can fix
these counterterms in a nonrenormalizable theory such as Einstein +
Maxwell \cite{DvN} because they cannot be present on the fundamental
level. They represent our ignorance about the ultraviolet completion
of gravity and their appearance is one of the inevitable limitations
of effective field theory \cite{BB,Donoghue}.

The reliable and significant parts of expression
(\ref{potentials:ratio charge}) are the factors of $1/3(a H x)^2$
and $\ln(a H x)$. The first of these is just the de Sitter
descendant of the short distance enhancement that was found for flat
space background \cite{LW2,BB}. Its presence represents a nice
correspondence check. The new, de Sitter feature is the enhancement
factor of $\ln(a H x)$. Both features are plotted in
Fig.~\ref{fig:point charge potential}.
\begin{figure}[ht]
\vskip -0.2cm
\includegraphics[scale=1.2]{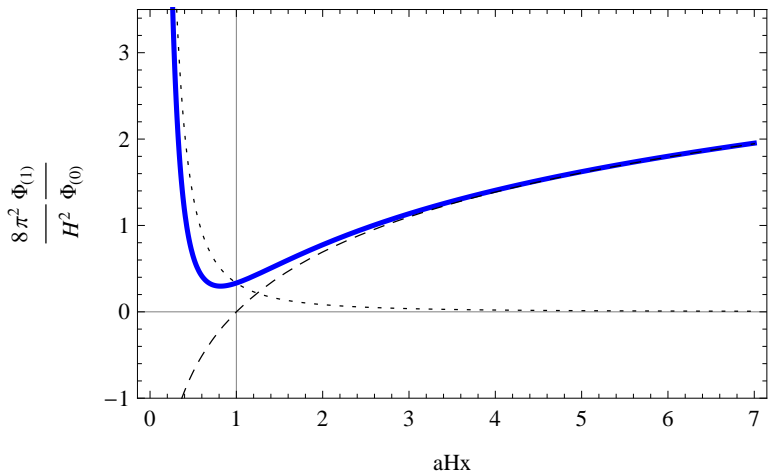}
\vskip -0.4cm \caption{\tiny The physically significant part of the
ratio $\Phi_{(1)}/\Phi_{(0)}$ in units of $\kappa^2 H^2/8\pi^2$, as
a function of the physical distance in Hubble units, $a H x$. The
solid blue curve gives $1/3a^2H^2a^2 + \ln(a H x)$. At short
distances the ratio is dominated by $1/3a^2H^2x^2$ (the short-dashed
line). At large physical distances the ratio is dominated by $\ln(a
H x)$ (the long-dashed line). \label{fig:point charge potential} }
\end{figure}

When viewed at fixed $x$, the distinctively de Sitter factor of
$\ln(a H x)$ in (\ref{potentials:ratio charge}) grows linearly in
the co-moving time $\ln(a) = H t$. Because this secular enhancement
factor multiplies the classical potential $\Phi_{(0)}$ it seems
reasonable to regard the effect as a time-dependent renormalization
of the source charge $q$, which contradicts the claim of Kitamoto
and Kitazawa that infrared gravitons screen gauge couplings
\cite{HKYK}. The slope is quite small. If one assumes single-scalar
inflation, the measured value of the scalar power spectrum and the
current limit on the tensor-to-scalar ratio \cite{SPT} imply
$\kappa^2 H^2/8 \pi^2 \leq 3.3 \times 10^{-11}$. Nevertheless, the
enhancement might be significant over a prolonged period of
inflation. Of course one is limited by the reliability of
perturbation theory; we cannot necessarily conclude that the
effective charge grows past the time at which $\kappa^2 H^2/8\pi^2
\times Ht \sim 1$ because the higher loop contributions reach the
same strength at this time. The reliable conclusion is rather that
perturbation theory breaks down; one must employ some kind of
nonperturbative resummation scheme to work out what happens later,
for example \cite{PTW}.

The secular enhancement cannot be understood in the same terms as
the screening from scalar quantum electrodynamics \cite{SQED,DW}
because both photon and graviton lines in the diagram of
Fig.~\ref{fig:photon} are uncharged. The explanation seems to derive
rather from the interpretation of force fields as transferring
momentum by the exchange of virtual particles. The typical
inflationary graviton carries a physical momentum of about $H$.
There is little effect near the source because the virtual photons
in this region carry much larger momenta. However, beyond a physical
Hubble distance the momenta of inflationary gravitons is larger so
scattering with them can give a virtual photon significantly more
momentum than it would otherwise carry.

To understand how an isotropic ensemble of gravitons can still
provide a net outward-directed push, consider the process in one
spatial dimension. Suppose we add a random momentum $\Delta p$ to
some fixed momentum $p > 0$,
\begin{equation}
p' = p + \Delta p \; .
\end{equation}
Even if the distribution of $\Delta p$ is symmetric about the origin
the distribution of the magnitude of $p'$ will still be asymmetric
about $\vert p' \vert = p$ because $\vert p'\vert > p$ receives
contributions from both $\Delta p > 0$ and $\Delta p < -2p$. In the
latter case the scattered virtual photon has $p' < -p$, so its
momentum is delivered to the direction opposite from which it
originally set out, but the force is still directed outward and
stronger than without the scattering. The $\ln(a) = Ht$ growth is
the same ``drunkard's walk'' factor as the magnitude of a massless,
minimally coupled scalar \cite{VFLS}, which also receives stochastic
accretions as successive modes experience horizon crossing.

These considerations are supported by the fields perceived by a an
observer who is held at a fixed physical distance from the source.
In Appendix~A we transform the field strength (\ref{F1i0}) to the
frame of this observer. Expressing the result (\ref{tilde F0i}) in
terms of a one loop correction to the scalar potential gives,
\begin{equation}
\widetilde{\Phi}_{(1)} = \widetilde{\Phi}_{(0)} \times
\frac{\kappa^2 H^2}{8 \pi^2} \Biggl\{ \frac{1}{3 H^2 r^2} + \ln(Hr)
+ \alpha
             - 2 \ln\left(\frac{a-1+Hr}{a-1-Hr}\right) \Biggr\}
\ ,
\label{tilde Phi}
\end{equation}
As before, only the first two terms are reliable and
significant.\footnote{\tiny The last term in~(\ref{tilde Phi}) is
negligible when $a\gg 1$ (since it gets suppressed as $1/a$) and the
third term can be removed by a suitable choice of the counterterm.}
However, both of these terms are constant, and the flat space factor
$1/3 H^2 r^2$ dominates the de Sitter correction $\ln(H r)$ for $H r
< 1$. That is just what one would expect because the typical
inflationary gravitons responsible for the secular growth of
(\ref{potentials:ratio charge}) have Hubble-scale physical momenta.
Only if one continues (\ref{tilde Phi}) to large values of $H r$ is
the logarithmic enhancement apparent.

\subsection{Point magnetic dipole} \label{dipole}

The current representing a point magnetic dipole of a strength $\vec m$ is
\begin{equation}
J^0(\eta,\vec{x}\,) = 0 \ , \ \ \ J^i(\eta,\vec{x}\,) =
-\epsilon^{ijk}m_j \partial_k \delta^3(\vec{x}\,) \; .
\end{equation}
The classical field strength tensor associated with it is,
\begin{equation}
F^{0i}_{(0)}(\eta,\vec{x}) = 0 \ , \ \ \ F^{ij}_{(0)}(\eta,\vec{x})
= \epsilon^{ijk} \Bigl( m_k\nabla^2 - \vec{m} \!\cdot\! \vec{\nabla}
\partial_k \Bigr) \, \frac1{4 \pi x} \; .
\end{equation}
Of course this system is described by a static and transverse vector
potential of the form (\ref{staticA1}), with $\mathcal{A}(\vec{x}) =
1/4\pi x$. Recall that the one loop field strengths for this case
are based on the intermediate quantity $\mathcal{J}_{
\mathcal{A}}(x)$ defined in expression (\ref{Jdef}). We can express
it in terms of derivatives acting on the integrals
$\mathcal{I}_{ab}$ given in Appendix B,
\begin{eqnarray}
\lefteqn{\kappa^2 \mathcal{J}_{\mathcal{A}}(\eta,x) \equiv - \!\!
\int \!\! d^4x' \, i \Bigl[F_{(1)}(x;x') \!+\! G_{(1)}(x;x')\Bigr]
\frac{\kappa^2}{4 \pi \Vert \vec{x}'\Vert } = -
\frac{\kappa^2H^2}{8\pi^2} \Bigl[\frac{1}{3} \ln(a) \!-\! \alpha
\!-\! \gamma \Bigr] \frac{1}{4\pi x} } \nonumber \\
& & \hspace{2.5cm} - \frac{a^{-1} \kappa^2 \partial^6}{1536 \pi^4}
\Bigl(\mathcal{I}_{22} \!-\! \mathcal{I}_{21}\Bigr) - \frac{\kappa^2
H^2}{128 \pi^4} \Biggl[ \Bigl(\frac{1}{12} \partial^4 \!-\!
\partial_0^2 \partial^2 \Bigr) \mathcal{I}_{12} \!-\!
\Bigl(\frac{1}{12} \partial^4 \!+\! \partial_0^2 \partial^2 \Bigr)
\mathcal{I}_{11} \Biggr] \; . \qquad
\end{eqnarray}
The various derivatives are acted in Appendix~C to produce the
result,
\begin{eqnarray}
\lefteqn{\kappa^2 \mathcal{J}_{\mathcal{A}}(\eta,x) = \frac{\kappa^2
H^2}{8 \pi^2} \Biggl\{ \frac{\theta(\Delta\eta_0 \!-\! x)}{4\pi x}
\left[\frac{1}{3 a^2 H^2 x^2} \!-\! \frac{1}{3} \ln(a H x) \!+\!
\alpha\!+\! \gamma \!-\!2\ln\left( \frac{1\!-\!a^{-1}\!+\!Hx}
{1\!-\!a^{-1}\!-\!Hx} \right) \right] } \label{Fij:magnetic} \nonumber \\
& & \hspace{2cm} + \frac{\theta(x \!-\! \Delta\eta_0)}{4\pi x}
\left[-\frac{1}{3} \ln(a) \!+\! \alpha \!+\! \gamma \!-\! 4 \!-\!
\frac{13}{3} \ln\left(\! 1 \!-\! \frac{1}{a} \right) \!-\!
\frac{1}{3(a\!-\!1)} \!-\! \frac{a}{3(a\!-\!1)^2} \right] \Bigg\} ,
\qquad
\end{eqnarray}
where $x= \| \vec{x} \|$ and $\Delta\eta_0 = \eta + 1/H$.

All the comments we made after equation (\ref{potentials:ratio
charge}) apply as well to (\ref{Fij:magnetic}). In particular, the
branch with $x > \Delta \eta_0$, which is not causally related to
the point source on the initial value surface, is nonsense based on
our having failed to perturbatively correct the initial state. Most
of those terms also fall off like powers of $1/a$. Even in the
causal branch with $x < \Delta \eta_0$, the out-going and in-coming
spherical wave is another artifact of the initial state, while the
factors of $\alpha$ and $\gamma$ derive from the finite parts of
higher derivative counterterms which parameterize our ignorance
about the true ultraviolet completion of Einstein + Maxwell. As
before, the terms which can be reliably fixed by low energy effective
field theory are just the factors of $1/3a^2H^2x^2$ and $-\frac13
\ln(a H x)$ on the first line of (\ref{Fij:magnetic}).

In the interests of simplicity we have used only the causal branch
to compute the quantity $\nabla^2/\partial^2 \,
\mathcal{J}_{\mathcal{A}}$ which determines the one loop field
strengths through relation (\ref{staticA2}),
\begin{eqnarray}
\lefteqn{ \frac{\nabla^2}{\partial^2} \, \kappa^2
\mathcal{J}_{\mathcal{A}} = \frac1{4\pi x} \times \frac{\kappa^2
H^2}{8 \pi^2} \Biggl\{ \frac{1}{3 a^2 H^2 x^2} \!-\! \frac{2}{3}
\ln(H x) \!+\! \alpha \!+\! \gamma \!-\! \frac13 \ln\Bigl( \frac{a
H x}{1 \!+\! a H x}\Bigr) } \nonumber \\
& & \hspace{.5cm} + \frac58 \ln\Bigl( \frac{1 \!-\! a^{-1} \!-\! Hx}
{1 \!-\! a^{-1} \!+\! Hx} \Bigr) \!+\! \frac1{12} \ln\Bigl( \frac{1
\!+\! a^{-1} \!-\! Hx} {1 \!+\! a^{-1} \!+\! Hx} \Bigr) \!+\!
\frac{(-\frac32 \!+\! \frac2{a}) H x}{ (1 \!-\! \frac1{a})^2 \!-\!
H^2 x^2 } \!+\! \frac{ \frac13 (1 \!-\! \frac1{a}) H x}{ [(1 \!-\!
\frac1{a})^2 \!-\! H^2 x^2]^2 } \Biggr\} \; . \qquad \label{Dragen}
\end{eqnarray}
Including the acausal branch would not affect the factors of
$1/3a^2H^2x^2$ and $-\frac23 \ln(H x)$ which are the only reliable
and significant parts of (\ref{Dragen}). From expression
(\ref{staticA2}) we see that the classical plus quantum vector
potential is,
\begin{equation}
\vec{A}(\eta,\vec{x}) = \vec{m} \! \times \! \vec{\nabla} \Biggl\{
\frac1{4\pi x} \Biggl[ 1 + \frac{\kappa^2 H^2}{8 \pi^2} \Bigl(
\frac1{3 a^2 H^2 x^2} - \frac23 \ln(H x) + {\rm Irrelevant} \Bigr) +
O(\kappa^4) \Biggl] \Biggr\} \; . \label{dipoleA}
\end{equation}

Of course the factor of $1/3a^2H^2x^2$ had to appear in
(\ref{dipoleA}) to give the correct flat space limit \cite{LW2}. The
striking things about the intrinsically de Sitter correction
$-\frac23 \ln(H x)$, relative to the analogous correction to the
point charge potential (\ref{potentials:ratio charge}), are the
opposite sign and the absence of secular growth at fixed $x$. These
features mean that an observer at fixed $x$ perceives no secular
change in the strength of the dipole, but different fixed $x$
observers report a screening of the dipole at increasing distance.
It might be significant that the one loop corrections to the fields
of a magnetic dipole from scalar quantum electrodynamics also show
weaker time dependence than the corrections to the fields of a point
charge \cite{DW}, although the change in that case was from
exponential screening to only linear screening.

We can use the formulae of Appendix A to transform the total
(classical plus quantum) field strengths to the frame of an observer
at a fixed physical distance $r = a x$ from the source. We shall
only include the one loop terms from the factors of $1/3a^2H^2x^2$
and $-\frac23 \ln(Hx)$ in (\ref{dipoleA}),
\begin{eqnarray}
\lefteqn{\widetilde{F}_{0i}(\tau,\vec{r}) = e^{H \tau} \sqrt{1 \!-\!
H^2 r^2} \, \frac{ H \epsilon^{ijk} m^j \widehat{r}^k}{ 4\pi r^2}
\Biggl\{ 1 } \nonumber \\
& & \hspace{5cm} + \frac{\kappa^2 H^2}{8 \pi^2} \Biggl[ \frac{5}{H^2
r^2} + \frac23 \Bigl[2 \!-\! \ln\Bigl( \frac{Hr}{a}\Bigr) \Bigr] +
{\rm Irrelevant} \Biggr] + O(\kappa^4) \Biggr\} \; ,
\label{ShunPei1} \qquad \\
\lefteqn{\widetilde{F}_{ij}(\tau,\vec{r}) = \frac{e^{H \tau}}{
\sqrt{1 \!-\! H^2 r^2}} \, \frac{\epsilon^{ijk}}{4\pi r^3} \Biggl\{
m^k \!-\! 3 \vec{m} \!\cdot\! \widehat{r} \widehat{r}^k \Bigl[1
\!-\! \frac23 H^2 r^2 \Bigr] + \frac{\kappa^2 H^2}{8 \pi^2} \Biggl[
\frac{3 m^k \!-\! 5 \vec{m} \!\cdot\! \widehat{r} \widehat{r}^k}{H^2
r^2} } \nonumber \\
& & \hspace{0cm} - \frac23 m^k \Bigl[1 \!+\! \ln\Bigl(\frac{H
r}{a}\Bigr) \Bigr] + 2 \vec{m} \!\cdot\! \widehat{r} \widehat{r}^k
\Bigl[\frac23 \!+\! \frac23 H^2 r^2 \!+\! \Bigl(1 \!-\! \frac23 H^2 r^2
\Bigr) \ln\Bigl( \frac{H r}{a} \Bigr) \Bigr] + {\rm Irrel.} \Biggr]
+ O(\kappa^4) \Biggr\} \; . \qquad \label{ShunPei2}
\end{eqnarray}
(We remind the reader that the scale factor is $a = e^{H \tau}
\sqrt{1 - H^2 r^2}$ in static coordinates.) Much of the complication
in expression (\ref{ShunPei2}) is to make the magnetic field
transverse in static coordinates, however, one can see that the
classical fields experience a secular enhancement by the factor
$\kappa^2 H^2/8 \pi^2 \times \frac23 H \tau$. Of course this secular
growth at fixed $r = a x$ is just the static coordinate reflection
of the $\ln(H x)$ screening we found in the conformal coordinate
result (\ref{dipoleA}).

\section{Conclusion}

In this work we have studied how inflationary gravitons influence
the electromagnetic field strengths of a point charge and a point
magnetic dipole in de Sitter space. This was done by solving the
quantum-corrected Maxwell's equations~\eqref{genEFE} at one loop
order, using the vacuum polarization recently calculated in
Ref.~\cite{LW1}. Results were derived for two types of observers, one
at a fixed position in co-moving coordinates --- and hence being
pulled away from the source by the inflationary expansion --- and
one at a fixed physical distance from the source.

For a point charge the co-moving observer perceives a Coulomb
potential (\ref{potentials:ratio charge}) which seems to describe a
secular renormalization of the charge by the factor $\kappa^2
H^2/8\pi^2 \times H t$. Even though the loop counting parameter is
very small (the most recent data \cite{SPT} implies $\kappa^2 H^2/8
\pi^2 \leq 3.3 \times 10^{-11}$ if one assumes single-scalar
inflation) this effect might be significant for a very long period
of inflation. It also represents another entry in the growing list
of secular effects mediated by inflationary gravitons
\cite{MW1,LW1,PMTW}.

The potential (\ref{tilde Phi}) of our static observer manifests
only a logarithmic running of the charge in space by the factor
$\kappa^2 H^2/8 \pi^2 \times \ln(H r)$. The physical interpretation
of both effects seems to be the momentum added to the force-carrying
virtual photons by the ensemble of Hubble-scale gravitons ripped out
of the vacuum by inflation. If these results and this interpretation
stand up they would contradict the claim by Kitamoto and Kitazawa
that inflationary gravitons screen gauge coupling constants
\cite{HKYK}.

The effect of inflationary gravitons on magnetic sources is weaker,
just as was found in a recent study of the vacuum polarization from
charged inflationary scalars \cite{DW}. Our co-moving observer
perceives a vector potential (\ref{dipoleA}) which contains no
secular change in the dipole, although it is consistent with a
logarithmic screening in space by the factor of $\kappa^2 H^2/8\pi^2
\times -\frac23 \ln(H x)$. Of course the static observer at fixed $r
= a x$ perceives fields (\ref{ShunPei1}-\ref{ShunPei2}) which
manifest a secular enhancement of the classical results by the
factor $\kappa^2 H^2/8\pi^2 \times \frac23 H \tau$.

We should also comment on the gauge issue. The vacuum polarization
from charged matter fields is gauge independent at one loop because
it involves only matter field propagators. However,
Fig.~\ref{fig:photon} shows that the contribution from gravitons
involves potential gauge dependence from both the photon propagator
and from the graviton propagator. An explicit study of this was made
in flat space background, using the 3-parameter family of Poincar\'e
invariant gauges \cite{LW2}. Although the single photon gauge
parameter dropped out there was massive dependence upon the two
graviton gauge parameters. In fact the flat space structure function
takes the form of a universal function whose form is dictated by
dimensionality and Poincar\'e invariance, times an algebraic
function of the graviton gauge parameters which can take any value
on the real line \cite{LW2}!

Of course this same gauge dependence must be present in the de
Sitter vacuum polarization \cite{LW1}, otherwise it would not
possess the correct flat space limit. However, the most important
corrections are intrinsically de Sitter; that is, they carry factors
of $H^2$ which vanish in the flat space limit. It has been suggested
that factors of $H^2 \ln(a)$ might be independent of the gauge
\cite{issues}. They do have a clear physical origin in the continual
production of inflationary gravitons. And there is precedence for
this idea from the behavior of flat space Green's functions. These
are highly gauge dependent, but they can be combined to give the
gauge independent S-matrix. Fortunately, we are not reduced to
opining about the possibility of gauge independence: the technology
exists to {\it check it} \cite{TW2,PMTW2} and we have begun work on
the project.

\section*{Acknowledgements}

We are grateful to S.~Deser and K.~E.~Leonard for conversations on
this subject. This work is part of the D-ITP consortium, a program 
of the Netherlands Organisation for Scientific Research (NWO) that 
is funded by the Dutch Ministry of Education, Culture and Science 
(OCW), it was partially supported by NWO Veni Project 
\# 680-47-406 by NSF grant PHY-1205591, and by the Institute for
Fundamental Theory at the University of Florida.

\section*{Appendix~A - Static coordinates on de Sitter space}

The transformation from conformal coordinates to static coordinates
$\widetilde{x}^\mu=(\tau,\vec{r}\,)$ is,
\begin{equation}
\tau = -\frac{1}{H} \ln\left[ \frac{1}{a(\eta)}\sqrt{1-H^2r^2}
\right] \ ,\ \ \ r^i= a(\eta) x^i \ ,
\end{equation}
where $a(\eta)=-1/H\eta$. The ranges are $-\infty< \tau <\infty$,
$0\le r=\|\vec{r}\,\|<{1}/{H}$, and the invariant line element in
these coordinates is,
\begin{equation}
ds^2 = -(1-H^2r^2) d\tau^2 + \frac{dr^2}{1-H^2r^2} + r^2d\Omega^2 \
,
\end{equation}
where $d\Omega^2$ is the line element squared on the unit 2-sphere.
In these coordinates an observer at $r$ is at a constant physical
distance from the origin, where our point sources are placed.

The electromagnetic field strength tensor in static coordinates is
given in terms of one in conformal coordinates as~\cite{DW},
\begin{align}
\widetilde{F}_{0i} ={}& \frac{e^{-2H \tau}}{1-H^2r^2} \Big{[}
-F^{0i} - H F^{ij} r^j \Big{]} \ , \label{tilde F0i} \\
\widetilde{F}_{ij} ={}& \frac{e^{-2H \tau}}{(1-H^2r^2)^2} \Big{[}
(1-H^2r^2) F^{ij} - 2 H^2 r^k F^{k [i} r^{j]} - 2 H F^{0[i} r^{j]}
\Big{]} \ , \label{tilde Fij}
\end{align}
where indices enclosed in square brackets are anti-symmetrized. Upon
inserting~(\ref{F1i0}) into Eq.~(\ref{tilde F0i}), one obtains (for
a late time observer within the Hubble distance, $Hr<{\rm
min}[1,a-1]$)
\begin{equation}
\widetilde{F}_{0i}(\vec{r}\,) = -\frac{q}{4\pi}
            \tilde\partial_i\left\{\frac{1}{r} +  \frac{\kappa^2H^2}{8\pi^2 r}
       \left[\frac{1}{H^2r^2} + \ln(Hr) + \alpha
             - 2\ln\left( \frac{a-1+Hr}{a-1-Hr}\right) \right] \right\}
   + {\cal O}(\kappa^4)
\ ,
\label{tilde F0i:2}
\end{equation}
where (covariant components of) the classical electric field is,
$\widetilde{F}_{0i}=-\tilde\partial_i [q/(4\pi r)]
=qr^i/(4\pi r^3)$.

\section*{Appendix~B - Integrals from section IV}

The four integrals appearing in section IV are
\begin{equation}
\mathcal{I}_{ab} = \int\limits_{-1/H}^{\eta} \! d\eta' \int \! d^3x' \,
  \frac{1}{\|\vec{x}'\|}
  \theta\big(\eta\!-\!\eta'\!-\!\|\vec{x}\!-\!\vec{x}^{\,\prime}\| \big)
            f_a(\eta',\vec{x}') g_b(\eta',\vec{x}') \ ,
\quad (a,b=1,2)
\,,
\end{equation}
where
\begin{equation}
f_1 = 1 \ , \ \ \ f_2 = -H\eta' \ , \ \ \ g_1 = 1 \ , \ \ \
g_2 = \ln \left[\frac{H^2}{4} \Big{(} (\eta\!-\!\eta')^2 \!-\! \|
        \vec{x}\!-\!\vec{x}^{\,\prime} \|^2 \Big{)} \right] \ .
\end{equation}
All the integrals are elementary, and they turn out to be
\begin{align}
\mathcal{I}_{11} = {}& \theta(\Delta\eta_0-x)\, 4\pi \Big{[}
\frac{x^3}{12} - \frac{x^2\Delta\eta_0}{6} +
\frac{\Delta\eta_0^3}{6} \Big{]} + \theta(x-\Delta\eta_0)\, 4\pi
\frac{\Delta\eta_0^4}{12x} \ , \label{I11}
\\
\mathcal{I}_{21} = {}& -H\eta\, \mathcal{I}_{11} +
\theta(\Delta\eta_0-x)\, 4\pi H \left[ \frac{\Delta\eta_0^4}{8} -
\frac{\Delta\eta_0^2x^2}{12} + \frac{x^4}{40} \right] +
\theta(x-\Delta\eta_0)\, 4\pi H \frac{\Delta\eta_0^4 
[4 \Delta\eta_0 \!-\! 5\eta]}{60x} \ , \label{I21}
\end{align}
\begin{align}
\mathcal{I}_{12}
 = {}& \theta(\Delta\eta_0-x)\, 4\pi
  \Bigg{\{} -\frac{1}{12x}(\Delta\eta_0^2-x^2)(\Delta\eta_0-x)^2
       \ln\Big[\frac{H}{2} (\Delta\eta_0-x) \Big]
\nonumber \\
& + \frac{1}{12x}(\Delta\eta_0^2 \!-\! x^2)(\Delta\eta_0 \!+\! x)^2
       \ln\Big[\frac{H}{2} (\Delta\eta_0 \!+\! x) \Big]
          \!+\! \frac{x^3}{6}\ln(Hx) \!-\! \frac{19}{72}x^3
          \!-\! \frac{4}{9}\Delta\eta_0(\Delta\eta_0^2 \!-\! x^2)  \Bigg{\}}
\nonumber \\
& + \theta(x-\Delta\eta_0) \, 4\pi
\left\{ \frac{\Delta\eta_0^4}{6x} \ln(H\Delta\eta_0)
- \frac{19}{72} \frac{\Delta\eta_0^4}{x} \right\}
\ ,
\label{I12}
\end{align}
\begin{align}
\mathcal{I}_{22}
={}& -H\eta\, \mathcal{I}_{12} + \theta(\Delta\eta_0-x)\, 4\pi H
\Bigg{\{} \left[ \frac{\Delta\eta_0^5}{15x} + \frac{\Delta\eta_0^4}{8}
     - \frac{\Delta\eta_0^2x^2}{12} + \frac{x^4}{40} \right]
            \ln\left[\frac{H}{2}(\Delta\eta_0+x) \right]
\nonumber \\
& + \left[ -\frac{\Delta\eta_0^5}{15x} + \frac{\Delta\eta_0^4}{8}
 - \frac{\Delta\eta_0^2x^2}{12} + \frac{x^4}{40} \right]
               \ln\left[ \frac{H}{2}(\Delta\eta_0-x) \right]
\!-\! \frac{77\Delta\eta_0^4}{240} \!+\!
\frac{59}{360}\Delta\eta_0^2x^2
    \!-\! \frac{19}{400}x^4 \Bigg{\}}
\nonumber \\
& + \theta(x-\Delta\eta_0)\, 4\pi
\left\{ \frac{2}{15} \frac{\Delta\eta_0^5}{x}
   \ln(H\Delta\eta_0) - \frac{46}{225} \frac{\Delta\eta_0^5}{x} \right\}
\ ,
\label{I22}
\end{align}
where $x=\| \vec{x} \|$ and $\Delta\eta_0=\eta+1/H$.

\section*{Appendix~C - Derivatives}

It is often necessary to act the flat space d'Alembertian on a
function of the conformal $\eta$ and just the magnitude $x \equiv
\Vert \vec{x} \Vert$ of the position vector. For this case we can
write,
\begin{equation}
\partial^2 f(\eta,x) = \frac1{x} \, (\partial_x \!-\! \partial_0)
(\partial_x \!+\! \partial_0) \Bigl[ x f(\eta,x) \Bigr] \; .
\end{equation}
This form is particularly effective when acting on out-going or
in-coming spherical waves: $f(x - \eta)/x$ or $f(x + \eta)/x$.

The rest is straightforward but tedious. The various derivatives of
section IV are,
\begin{align}
\partial^4\, \mathcal{I}_{11} ={}& \frac{8\pi}{x}
\ ,
\\
\partial_0^2\partial^2 \, \mathcal{I}_{11}
={}& - \theta(x-\Delta\eta_0)\, \frac{8\pi}{x}
\ ,
\\
\partial^4\, \mathcal{I}_{12}
 ={}& \theta(\Delta\eta_0-x) \frac{8\pi}{x}
       \Big{[} 2\ln(Hx) + 1 \Big{]}
 + \theta(x-\Delta\eta_0) \frac{8\pi}{x}
    \left[ 2\ln\left( 1 - \frac{1}{a} \right) + 1 \right]
\ ,
\\
\partial_0^2\partial^2 \mathcal{I}_{12}
 ={}& \theta(\Delta\eta_0-x)\, \frac{8\pi}{x}
 \ln\left[ \frac{1-a^{-1}-Hx}{1-a^{-1}+Hx} \right]
- \theta(x-\Delta\eta_0)\, \frac{8\pi}{x}
 \left[2 \ln\left( 1-\frac{1}{a} \right) + 1 \right]
\ ,
\\
\partial^6\, \mathcal{I}_{21} ={}& 0
\ ,
\\
\partial^6\, \mathcal{I}_{22}
  ={}& - \theta(\Delta\eta_0-x)\, \frac{16\pi}{a x^3}
+ \theta(x-\Delta\eta_0) \, \frac{16\pi H^2}{x}
      \left[ \frac{a^2}{(a-1)^2} + \frac{a}{a-1} \right]
\ .
\end{align}
Note that no delta functions appear from taking derivatives of step
functions in~(\ref{I11}--\ref{I22}) since the coefficients
multiplying them are zero.

\end{document}